%% file: main.tex
\documentclass{Interspeech2024}

\interspeechcameraready

\usepackage{amsmath,graphicx}

\usepackage{graphicx}
\usepackage{amssymb, amsmath, bm, mathtools,array}
\usepackage{textcomp}
\usepackage{hyperref}
\usepackage{verbatim,lipsum}
\usepackage{booktabs}
\usepackage{tcolorbox}
\usepackage{lscape}
\usepackage{subcaption}
\usepackage{multirow}
\usepackage{colortbl}
\usepackage{color}
\usepackage{multibib}
\usepackage{tikz}
\usetikzlibrary{positioning}
\usetikzlibrary{shapes}
\newcommand{\ideaname}{compositional data analysis}
\newcommand{\symbolset}[1]{\boldsymbol{\mathrm{#1}}}
\newcommand{\symbolvec}[1]{\boldsymbol{#1}}
\newcommand{\symbolhp}{{H}}
\newcommand{\symbolirled}[1]{\tilde{#1}}

\newcommand{\scoreasv}{s_{\text{asv}}}
\newcommand{\scorecm}{s_{\text{cm}}}
\newcommand{\scoresasv}{s_{\text{sasv}}}
\newcommand{\cllr}{\mathrm{Cllr}}

\newcommand{\classtar}{\mathsf{tar.bf}}
\newcommand{\classnon}{\mathsf{non.bf}}
\newcommand{\classspoof}{\mathsf{spf}}

\newcommand{\classasvp}{\mathsf{tar}}
\newcommand{\classasvn}{\mathsf{non}}
\newcommand{\classcmp}{\mathsf{bona}}
\newcommand{\classcmn}{\mathsf{spoof}}
\newcommand{\evidence}{\boldsymbol{x}}
\newcommand{\evidencenew}[1]{\boldsymbol{#1}}
\newcommand{\evidenceset}{\symbolset{X}}

\newcommand{\sysbase}{{\texttt{B1}}}
\newcommand{\sysbasec}{{\texttt{B1c}}}
\newcommand{\syssi}{{\texttt{L2}}}
\newcommand{\syssic}{{\texttt{L2c}}}
\newcommand{\syssii}{{\texttt{L3}}}
\newcommand{\syssiic}{{\texttt{L3c}}}
\newcommand{\syssrefi}{{\texttt{B1v2}}}
\newcommand{\syssrefii}{{\texttt{Post}}}

\newcommand{\irli}{\symbolirled{w}_1}
\newcommand{\irlii}{\symbolirled{w}_2}

\newcommand{\llr}[3]{\mathsf{llr}^{#2}_{#3}(#1)}

\DeclareMathOperator{\ilr}{ilr}

\title{Revisiting and Improving Scoring Fusion for Spoofing-aware Speaker Verification Using Compositional Data Analysis}

\name[affiliation={1}]{Xin}{Wang}
\name[affiliation={2}]{Tomi}{Kinnunen}
\name[affiliation={3}]{Kong Aik}{Lee}
\name[affiliation={4}]{Paul-Gauthier}{Noé}
\name[affiliation={1}]{Junichi}{Yamagishi}

\address{
  $^1$National Institute of Informatics, Japan,
  $^2$School of Computing, University of Eastern Finland, Finland, 
  $^3$Department of Electrical and Electronic Engineering, The Hong Kong Polytechnic University, Hong Kong, 
  $^4$Laboratoire Informatique d’Avignon, Avignon Université, France}
\email{wangxin@nii.ac.jp}


\begin{document}
\input{note}
\maketitle

\begin{abstract}
Fusing outputs from automatic speaker verification (ASV) and spoofing countermeasure (CM) is expected to make an integrated system robust to zero-effort imposters and synthesized spoofing attacks. Many score-level fusion methods have been proposed, but many remain heuristic. This paper revisits score-level fusion using tools from decision theory and presents three main findings. First, fusion by summing the ASV and CM scores can be interpreted on the basis of compositional data analysis, and score calibration before fusion is essential. Second, the interpretation leads to an improved fusion method that linearly combines the log-likelihood ratios of ASV and CM. However, as the third finding reveals, this linear combination is inferior to a non-linear one in making optimal decisions. The outcomes of these findings, namely, the score calibration before fusion, improved linear fusion, and better non-linear fusion, were found to be effective on the SASV challenge database.
\end{abstract}

\section{Introduction}
Automatic speaker verification (ASV) systems are vulnerable to spoofed synthetic speech data that clone the target speakers' voices \cite{Wu2015}. 
By combining ASV and spoofing countermeasure (CM), which ideally should reject any spoofed data, we expect to be able to construct a spoofing-aware ASV (SASV) system that only accepts speech data from real (bona fide) humans who match the target speaker's identity. 

An SASV system can be a cascade of ASV and CM sub-systems \cite{hadid2015biometrics}. 
The ASV and CM make separate binary decisions, and the input is accepted if it is accepted by both sub-systems.
Another design used by many SASV systems is to produce one score and make one binary decision. 
Let the input be $\evidence =(\evidence^{(p)}, \evidence^{(r)}) \in\mathcal{X}$, where $\evidence^{(p)}$ and $\evidence^{(r)}$ are the probe (test) and reference (enrollment) samples, respectively, and can be waveforms, features, etc. 
An SASV system defines a scoring function $g: \mathcal{X}\rightarrow\mathbb{R}$ that maps $\evidence$ into a score $\scoresasv$. If $\scoresasv$ surpasses a threshold, the system takes the action of $\mathtt{accept}$, claiming that $\evidence_{p}$ is bona fide \emph{and} matches the speaker identity in $\evidence_{r}$. Otherwise, it takes the action of $\mathtt{reject}$. 

This paper focuses on the SASV systems that produce $\scoresasv$ using score-level fusion. The idea is to use ASV and CM sub-systems to obtain an ASV score $\scoreasv$ and CM score $\scorecm$ then use a score-level fusion function $h:\mathbb{R}\times\mathbb{R}\rightarrow\mathbb{R}$ to merge the scores into $\scoresasv$. 
Note that there are feature-level fusion and end-to-end methods \cite{sizov2015joint,gomez-alanisJoint2021, jung22c_interspeech}, which do not need to produce $\{\scoreasv, \scorecm\}$. We argue that score-level fusion is useful because $\{\scoreasv, \scorecm\}$ provides invaluable extra explanations on the model decision. Some evaluation metrics, such as tandem equal error rate (t-EER) \cite{kinnunenTEER2023}, also require separated $\scoreasv$ and $\scorecm$. 

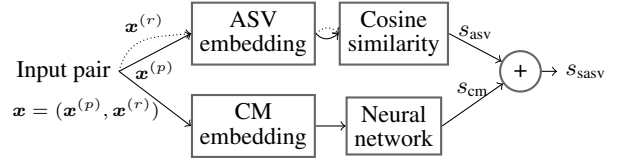
\begin{figure}[!t]
\begin{center}

\begin{tikzpicture}[
data/.style={},
block/.style={rectangle, thick, draw=black!60},
operation/.style={circle, thick, draw=black!60},
line/.style={black, ->},
annotation/.style={midway,above},
]
\node[data] (input) at (0, 1)   {\shortstack{Input pair} }; 
\node[data] (name) at (0.3, 0.5)   {{\scriptsize $\evidence=(\evidence^{(p)}, \evidence^{(r)})$}}; 
\node[block] (asv) at (2.5, 1.5)   {\shortstack{ASV \\ embedding }}; 
\node[block] (cm) [below=0.8 of asv.center] {\shortstack{CM \\ embedding }}; 
\node[block] (s_asv) [right=0.3 of asv] {\shortstack{Cosine \\ similarity}}; 
\node[block] (s_cm) at (cm -| s_asv) {\shortstack{Neural \\ network}}; 
\node[operation] (sum) [right=5.0 of input] {+}; 
\node[data] (s_sasv) [right=0.2 of sum] {$\scoresasv$}; 

\draw[line] (input.east) -- node[annotation, below]{ {\scriptsize $\evidence^{(p)}$} } (asv.west);
\draw[line, densely dotted] (input.east) to [out=75,in=185,looseness=1.2]  node[annotation]{ {\scriptsize $\evidence^{(r)}$} } (asv.west);
\draw[line] (input.east) -- (cm.west);
\draw[line] (asv.east) to [out=315,in=215,looseness=1.2] (s_asv.west);
\draw[line, densely dotted] (asv.east) to [out=45,in=135,looseness=1.2] (s_asv.west);
\draw[line] (cm) -- (s_cm);
\draw[line] (s_asv.east) -- node[annotation]{$\scoreasv$} (sum);
\draw[line] (s_cm.east) -- node[annotation]{$\scorecm$} (sum);
\draw[line] (sum) -- (s_sasv);
\end{tikzpicture}
\vspace{-6mm}
\caption{Example SASV with score-level fusion (B1 in \cite{jung22c_interspeech})}
\vspace{-6mm}
\label{fig:sasv}
\end{center}
\end{figure}

An example SASV system using score-level fusion (baseline B1 of the SASV challenge \cite{jung22c_interspeech}) is illustrated in Fig.~\ref{fig:sasv}. 
Its ASV sub-system extracts embeddings from $\evidence^{(p)}$ and $\evidence^{(r)}$ and computes their cosine similarity as $\scoreasv$, which indicates the degree that $\evidence^{(p)}$ and $\evidence^{(r)}$ match in terms of speaker identity. 
The CM sub-system extracts an embedding from $\evidence^{(p)}$ and produces a score $\scorecm\in\mathbb{R}$ using a neural network. A higher $\scorecm$ indicates that $\evidence^{(p)}$ is more likely to be bona fide. Hence, it is intuitively reasonable to fuse the scores by $\scoresasv =  \scoreasv + \scorecm$.  

More advanced fusion functions have been proposed in the SASV and other biometrics fields \cite{chingovskaAntispoofing2013, hadid2015biometrics}. 
Although most are intuitively reasonable, for example, normalizing the numeric range of $\scoreasv$ and $\scorecm$ before summation \cite{jung22c_interspeech}, the link between practice and theory is not always clear. Therefore, the implicit assumptions, link to optimal decisions, and room for improvement remain elusive. 

Starting from the summation of the ASV and CM scores, we scrutinize the score fusion using tools from decision theory.  
After reformulating SASV as a classification task that involves three data classes but requires a binary decision,
we interpret SASV scoring from the perspective of \ideaname{} and show that the summation of  $\scoreasv$ and $\scorecm$ is a reasonable choice 
if the ASV and CM scores are proper log-likelihood ratios (LLRs).
Hence, the interpretation indicates that score calibration before fusion is beneficial. 
The interpretation also leads to an improved fusion method that uses probabilistic models to compute LLRs for summation. Finally, the summation-based fusion is analyzed from the perspective of optimal decision policy and found to be theoretically inferior to a non-linear fusion method.

The above interpretations of SASV have not been presented as far as the authors are aware. They differ from one existing study that treats $\{\scoreasv, \scorecm\}$ as posterior probabilities \cite{zhangProbabilistic2022}. Experiments were conducted on the SASV challenge database \cite{jung22c_interspeech}. The calibration before fusion was found to be effective, especially for linear fusion by summation. The linear and non-linear fusion of LLRs outperformed baselines, and the latter outperformed the systems fusing posterior probabilities \cite{zhangProbabilistic2022}. The results are reproducible. 
\footnote{Code site: https://github.com/nii-yamagishilab/SpeechSPC-mini.}

\section{Interpreting and improving SASV score fusion}
\label{sec:theory}

\subsection{SASV is not fusion of congruent binary classifiers}
\label{sec:theoryold}
By definition, an SASV system should only $\mathtt{accept}$ if the input is uttered by the bona fide target speaker; otherwise, $\mathtt{reject}$.
Fusing ASV and CM scores for a binary decision in the SASV system seems to be similar to the score fusion in multi-modal biometric verification \cite{nandakumar2007likelihood, morrison2013tutorial}. 

Let $\evidenceset{} = \{\evidence{}_1, \cdots, \evidence{}_K\}$ be the inputs to a biometric verification system with $K$ modalities (e.g., speech and face). The system determines whether inputs match the claimed identity ($\symbolhp{}_{\classasvp{}}$) or not ($\symbolhp{}_{\classasvn{}}$).  
Assuming that $\evidence_k$ are statistically independent, we can use the Bayes' formula \cite{dudaPattern2001} to obtain
\begin{equation}
\log\frac{{P(\symbolhp{}_{\classasvp{}} | \evidenceset{})}}{1-{P(\symbolhp{}_{\classasvp{}} | \evidenceset{})}} = \log\frac{\pi_{\classasvp{}}}{1-\pi_{\classasvp{}}} + \sum_{k=1}^{K} \llr{\evidence_k}{\classasvp{}}{\classasvn{}},
\label{eq:bayes}
\end{equation}
where 
$\llr{\evidence_k}{\classasvp{}}{\classasvn{}} \triangleq \log\frac{p( \evidence_k | \symbolhp{}_{\classasvp{}})}{p( \evidence_k | \symbolhp{}_{\classasvn{}})} $ is an LLR produced by the $k$-th sub-system, and $\pi_{\classasvp{}}$ is the prior probability of class $\classasvp{}$. 
Assuming that the LLRs and priors are equal to the ground truth,
our decision strategy for minimizing the decision error rate is to accept ${\classasvp{}}$ if and only if 
$\sum_{k=1}^{K} \llr{\evidence_k}{\classasvp{}}{\classasvn{}} + \log\frac{\pi_{\classasvp{}}}{1-\pi_{\classasvp{}}} > 0$  
\cite{dudaPattern2001}.
The sum of LLRs is the weight of evidence to favor $\classasvp{}$, which is known as \emph{independent additivity} of LLRs \cite{jaynes2003probability}.

We may hastily treat SASV as biometric verification with two modalities, namely, ASV and CM. Thus, fusion by $\scorecm + \scoreasv$ can be interpreted as the sum of LLRs, given that the scores approximate the LLRs. 
However, unlike Eq.~(\ref{eq:bayes}) where all sub-systems deal with $\{\symbolhp{}_{\classasvp{}}, \symbolhp{}_{\classasvn{}}\}$,
ASV and CM have different hypotheses: for CM, the input is either bona fide ($\symbolhp{}_{\classcmp{}}$) or spoofed ($\symbolhp{}_{\classcmn{}}$). 
With four hypotheses involved, independent additivity of LLRs and Eq.~(\ref{eq:bayes}) no longer hold \cite[sec.4.3]{jaynes2003probability}.

\subsection{Interpreting SASV using \ideaname{}} 
A better starting point is to have three exhaustive and mutually exclusive hypotheses \cite{hadid2015biometrics, chingovska2014biometrics}:  
the input probe is bona fide and matches the reference ($\symbolhp{}_{\classtar{}}$), bona fide and unmatched ($\symbolhp{}_{\classnon{}}$), or spoofed ($\symbolhp{}_{\classspoof{}}$). \footnote{It is unnecessary to separate $\symbolhp{}_{\classspoof{}}$ into matched and non-matched cases since both should be rejected \cite{kinnunenTEER2023}.} 
To interpret this classification task with three classes but two actions $\{\mathtt{accept}, \mathtt{reject}\}$, we use \emph{\ideaname{}} \cite{aitchison1982statistical,noe:tel-04264175}.

Let $\symbolvec{P} = [P(\symbolhp{}_{\classspoof{}} | \evidence), P(\symbolhp{}_{\classnon{}} | \evidence), P(\symbolhp{}_{\classtar{}} | \evidence)]^\top$ and $\boldsymbol{\pi} = [\pi_{\classspoof{}}, \pi_{\classnon{}}, \pi_{\classtar{}}]^\top$ be the vectors of posterior and prior probabilities, respectively. Let the vector of likelihoods be $\boldsymbol{w} = [p( \evidence | \symbolhp{}_{\classspoof{}} ), p( \evidence | \symbolhp{}_{\classnon{}}), p( \evidence | \symbolhp{}_{\classtar{}})]^\top$.
Because $\sum_i\pi_i = \sum_i P_i = 1$, where $i$ is the index of the vector's element, $\symbolvec{\pi}$ and $\symbolvec{P}$ are compositional vectors on the probability simplex $\mathbb{S}^3$. 
Hence, $\symbolvec{P}$ (and $\symbolvec{\pi}$) can be equivalently represented by a vector in 2D Euclidean space $\symbolirled{\symbolvec{P}} = [\symbolirled{{P}_1}, \symbolirled{{P}_2}]^\top\in\mathbb{R}^2$. 

Among many possible choices for $\mathbb{S}^3\rightarrow\mathbb{R}^2$, an isometric-log-ratio (ILR) transformation \cite{egozcue2003isometric} $\symbolirled{\symbolvec{P}} = \ilr(\symbolvec{P})$  defines  
\begin{align}
\symbolirled{{P}}_1 = \frac{1}{\sqrt{2}}\log\frac{P_2}{P_1}, \quad \symbolirled{{P}}_2 = \frac{1}{\sqrt{6}}\log\frac{P_3P_3}{P_1P_2},
\label{eq:p2}
\end{align}
where the scalars $\frac{1}{\sqrt{2}}$ and $\frac{1}{\sqrt{6}}$ are due to the Aitchison orthonormal basis \cite[Sec.10.2]{noe:tel-04264175}\cite{egozcue2003isometric}. The same transformation $\ilr(\cdot)$ can be applied to $\symbolvec{\pi}$ and $\symbolvec{w}$. Particularly, for $\symbolirled{\boldsymbol{w}}$, we have
\begin{equation}
\label{eq:w1}
\begin{split}
\symbolirled{{w}}_1 &= \frac{1}{\sqrt{2}}\log\frac{w_3w_2}{w_1w_3} = \frac{1}{\sqrt{2}}\Big(\llr{\evidence}{\classtar{}}{\classspoof{}} -\llr{\evidence}{\classtar{}}{\classnon} \Big),
\end{split}
\end{equation}
\begin{equation}
\label{eq:w2}
\begin{split}
\symbolirled{{w}}_2 &= \frac{1}{\sqrt{6}}\log\frac{w_3 w_3}{ w_1w_2} = \frac{1}{\sqrt{6}}\Big(\llr{\evidence}{\classtar{}}{\classnon} + \llr{\evidence}{\classtar{}}{\classspoof{}}  \Big).
\end{split}
\end{equation}

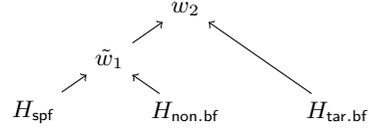
\begin{figure}[!t]
\begin{center}
\begin{tikzpicture}[
data/.style={},
line/.style={black, ->},
]

\node[data] (p0) at (0, 0)   {$\symbolhp_\classspoof$}; 
\node[data] (p1) at (2, 0)   {$\symbolhp_\classnon$}; 
\node[data] (p2) at (4, 0) {$\symbolhp_\classtar$}; 
\node[data] (r1) at (1, 0.7) {$\irli$}; 
\node[data] (r2) at (2, 1.4) {$\irlii$}; 
\draw[line] (p0) -- (r1);
\draw[line] (p1) -- (r1);
\draw[line] (r1) -- (r2);
\draw[line] (p2) -- (r2);
\end{tikzpicture}
\vspace{-3mm}
\caption{Bifurcating tree for ternary hypothesis test based on \ideaname{} \cite{noe:tel-04264175}}
\vspace{-5mm}
\label{fig:tree}
\end{center}
\end{figure}

With the $\symbolirled{\boldsymbol{P}}, \symbolirled{\boldsymbol{\pi}}$, and $ \symbolirled{\boldsymbol{w}}$, it is known that $\symbolirled{\boldsymbol{P}} = \symbolirled{\boldsymbol{\pi}} + \symbolirled{\boldsymbol{w}}$ \cite[Sec.3.3.3]{noe:tel-04264175}.
This equation has a similar form of ``posterior log-odds = prior log-odds + LLR'' to Eq.~(\ref{eq:bayes}). 
Similar to the sum of LLRs in Eq.~(\ref{eq:bayes}), $\symbolirled{\symbolvec{w}} = [\symbolirled{{w}_1}, \symbolirled{{w}_2}]^\top$ can be interpreted as weights of evidence, albeit in a hierarchical manner \cite{noe:tel-04264175} as illustrated in Fig.~\ref{fig:tree}. A larger $\symbolirled{w}_2$ favors $\symbolhp{}_{\classtar{}}$ over the other two, 
and it is the $\symbolirled{w}_1$ that discriminates  $\symbolhp{}_{\classspoof{}}$ from $\symbolhp{}_{\classnon{}}$.

The above description indicates that $\symbolirled{w}_2$ is what an SASV system needs to produce. This observation paves the way for interpreting and improving the SASV score fusion methods.

\subsection{Fusion method 1: summation of ASV and CM scores}
Note that ASV is designed to discriminate $\classtar{}$ and $\classnon{}$, while CM  differentiates $\classtar{}$ from $\classspoof{}$.
\textbf{If} their scores approximate the LLRs, i.e., $\scoreasv\approx\llr{\evidence}{\classtar{}}{\classnon}$ and $\scorecm\approx\llr{\evidence}{\classtar{}}{\classspoof{}}$,\footnote{For a speaker-independent CM, we can assume $p(\evidence | \classtar{})\approx p(\evidence | \classnon{}) \approx p(\evidence | \classcmp{})$ \cite{kinnunenTandem2020} and $\llr{\evidence}{\classtar{}}{\classspoof{}}\approx\llr{\evidence}{\classcmp{}}{\classspoof{}}$.} we can follow Eq.~(\ref{eq:w2}) and compute $\scoresasv$ as
\begin{equation}
\symbolirled{{w}}_2\approx\scoresasv = \frac{1}{\sqrt{6}} (\scoreasv + \scorecm).
\label{eq:fusion_score}
\end{equation}  
Hence, the summation-based score fusion is interpretable from the perspective of \ideaname{}. 
Note that the factor $\frac{1}{\sqrt{6}}$ does not affect the discrimination power of $\scoresasv$.

The interpretation may be disappointing since Eq.~(\ref{eq:fusion_score}) is the same as what we have done, but it suggests a practice that is either ignored or conducted arbitrarily: \emph{the ASV and CM scores need to well approximate the LLRs before summation.}
One way to do so is score calibration \cite{morrison2013tutorial}. 
For example, let  $\{f_{\text{asv}}, f_{\text{cm}}\}$ be calibration functions, both of which have a form of affine transformation  $f(x)=ax+b$. 
We can learn parameters of $\{f_{\text{asv}}, f_{\text{cm}}\}$ on a development set through logistic regression \cite{brummer13_interspeech}, after which we compute $ \scoresasv = \frac{1}{\sqrt{6}} \big[f_{\text{asv}}(\scoreasv) + f_{\text{cm}}(\scorecm)\big]$.

\subsection{Fusion method 2: summation of LLRs}
\label{sec:llrsum}  
Rather than calibrating the ASV and CM scores to produce the LLRs (i.e., discriminative calibration \cite{VanLeeuwen2014}), we can treat the scores as features and carry out generative calibration \cite{VanLeeuwen2014}. 
Let us create a feature vector $\evidencenew{s}=[\scoreasv, \scorecm]^{\top}$ and 
introduce likelihood functions $p(\evidencenew{s};  \boldsymbol{\theta}_\classtar{})$,  $p(\evidencenew{s}; \boldsymbol{\theta}_{\classspoof{}} )$, and $p( \evidencenew{s} ; \boldsymbol{\theta}_{\classnon{}})$ for the three classes.  Assuming $\evidencenew{s}$ encodes all the information of $\evidence$ and  $\frac{p(\evidence | \symbolhp{}_1)}{p(\evidence | \symbolhp{}_2)} = \frac{p(\evidencenew{s} | \boldsymbol{\theta}_1)}{p(\evidencenew{s} | \boldsymbol{\theta}_2)}$, based on Eq.~(\ref{eq:w2}), we can compute
\begin{equation}
\scoresasv = \frac{1}{\sqrt{6}}\Big(\llr{\evidencenew{s}}{\classtar{}}{\classnon} + \llr{\evidencenew{s}}{\classtar{}}{\classspoof{}}  \Big),
\label{eq:fusionthreegaussians}
\end{equation}
where $\llr{\evidencenew{s}}{\classtar{}}{\classnon}\triangleq \log\frac{p( \evidencenew{s} ; \boldsymbol{\theta}_{\classtar{}})}{p( \evidencenew{s}; \boldsymbol{\theta}_{\classnon{}})} $ and 
$\llr{\evidencenew{s}}{\classtar{}}{\classspoof}\triangleq \log\frac{p( \evidencenew{s} ; \boldsymbol{\theta}_{\classtar{}})}{p( \evidencenew{s}; \boldsymbol{\theta}_{\classspoof{}})}$.
Each $p(\evidencenew{s};  \boldsymbol{\theta})$ can be a Gaussian or another parametric distribution, and its parameter set $\boldsymbol{\theta}$
can be estimated by maximizing the likelihood on development data. 

During testing, we compute  $p(\evidencenew{s};  \boldsymbol{\theta}_\classtar{})$,  $p(\evidencenew{s}; \boldsymbol{\theta}_{\classspoof{}} )$, and $p( \evidencenew{s} ; \boldsymbol{\theta}_{\classnon{}})$ of a test sample and plug them into Eq.~(\ref{eq:fusionthreegaussians}). 
Since the selected parametric distributions may be different from the actual distribution of $\evidencenew{s}$, the estimated LLRs may be inaccurate. Hence, we can optionally apply discriminative calibration on the LLRs before fusion.

\subsection{Fusion method 3: nonlinear fusion of LLRs}
\label{sec:nonlinear}
The interpretation so far leads to the simple form of score fusion in Eqs.~(\ref{eq:fusion_score}) and (\ref{eq:fusionthreegaussians}). Importantly, however, it also reveals a major limitation for making decisions.
Recall from Sec.~\ref{sec:theoryold} that, for a single ASV system ($K=1$), we accept ${\classasvp{}}$ if and only if $\llr{\evidencenew{s}}{\classasvp{}}{\classasvn{}} + \log\frac{\pi_{\classasvp{}}}{1-\pi_{\classasvp{}}}>0$.
For SASV, a similar decision policy is to accept ${\classtar{}}$ if and only if $\symbolirled{{P}}_2 = \symbolirled{{\pi}}_2 + \symbolirled{{w}}_2 > 0$. Given $\symbolirled{{P}}_2$ defined in Eq.~(\ref{eq:p2}), this condition can be written as
\begin{equation}
P(\symbolhp{}_{\classtar{}} | \evidence  ) ^ 2 > P(\symbolhp{}_{\classnon{}} | \evidence  )P(\symbolhp{}_{\classspoof{}} | \evidence),
\label{eq:action_fused}
\end{equation} 
or equivalently, 
\begin{equation}
\begin{split}
\llr{\evidencenew{s}}{\classtar{}}{\classnon} + &\llr{\evidencenew{s}}{\classtar{}}{\classspoof{}} 
>
\log\frac{\pi_{\classspoof{}} \pi_{\classnon{}} }{\pi_{\classtar{}} ^ 2}.
\end{split}
\label{eq:action_fused_lr}
\end{equation} 
However, this decision policy cannot minimize the decision cost even if when the true LLRs and priors are given.

It can be shown that, for SASV with equal decision costs on false rejection of $\classtar{}$ and false acceptance of $\classnon$ or $\classspoof$, the optimal decision policy is to accept $\classtar{}$ when
\footnote{It is different from the condition $P(\symbolhp{}_{\classtar{}} | \evidence) > P(\symbolhp{}_{\classnon{}} | \evidence)$ AND $P(\symbolhp{}_{\classtar{}} | \evidence) >  P(\symbolhp{}_{\classspoof{}} | \evidence)$ \cite[sec.10.3]{noe:tel-04264175}, which maximizes the decision utility \cite{he2006three} with a scalar diagonal utility matrix.}
\begin{equation}
P(\symbolhp{}_{\classtar{}} | \evidence) > P(\symbolhp{}_{\classnon{}} | \evidence) + P(\symbolhp{}_{\classspoof{}} | \evidence),
\label{eq:action_optimal}
\end{equation}
or with Bayes' formula and logarithm applied,
\begin{equation}
-\log[(1-\rho) e^{-\llr{\evidencenew{s}}{\classtar{}}{\classnon}} + \rho e^{- \llr{\evidencenew{s}}{\classtar{}}{\classspoof{}}}] > -\log\beta,
\label{eq:action_optimal_llr}
\end{equation}
where $\beta \triangleq \frac{  \pi_{\classtar{}}}{\pi_{\classnon{}} + \pi_{\classspoof{}}}$ and $\rho \triangleq \frac{ \pi_{\classspoof{}} }{\pi_{\classnon{}} + \pi_{\classspoof{}}}$. \footnote{$\rho$ is the same as the spoof prevalence prior in t-EER \cite{kinnunenTEER2023}.}

Inequality.~(\ref{eq:action_fused}) is a necessary but not sufficient condition to achieve Ineq.~(\ref{eq:action_optimal}). 
For example, a sample with $[P(\symbolhp{}_{\classspoof{}} | \evidence), P(\symbolhp{}_{\classnon{}} | \evidence), P(\symbolhp{}_{\classtar{}} | \evidence)] = [0.05, 0.65, 0.3]$ will be rejected by Ineq.~(\ref{eq:action_optimal}) but falsely accepted by (\ref{eq:action_fused}). 
The overall decision cost cannot be lower than that based on the optimal decision policy in Ineq.~(\ref{eq:action_optimal}) (or (\ref{eq:action_optimal_llr})), given that we know the true LLRs and priors.

\begin{figure}[!t]
\begin{center}
\includegraphics[width=0.9\columnwidth]{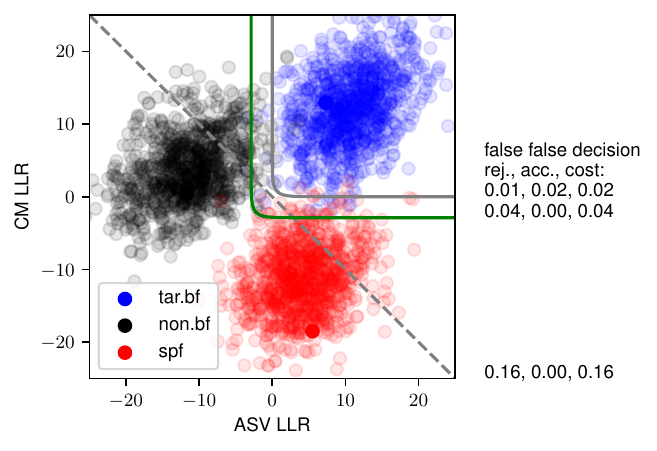}
\vspace{-3mm}
\caption{Scatter plot of ASV $\llr{\evidencenew{s}}{\classtar{}}{\classnon}$ and CM $\llr{\evidencenew{s}}{\classtar{}}{\classspoof{}}$ from simulated data. Dashed and solid lines are decision boundaries based on Ineqs.~(\ref{eq:action_fused_lr}) and (\ref{eq:action_optimal_llr}), respectively, given true flat prior. Green solid line is based on Ineq.~(\ref{eq:action_optimal_llr}) but mis-matched priors ${\pi}_\classspoof{}=0.05, {\pi}_\classnon{}=0.05, {\pi}_\classtar{}=0.9$. Note that LLRs rather than raw scores are plotted.}  
\vspace{-5mm}
\label{fig:simplex}
\end{center}
\end{figure}

To support the above argument, we generated ASV and CM scores using Gaussian distributions and a uniform prior, which enables us to compute the true LLRs and compared the linear and non-linear decision boundaries. 
As Fig.~\ref{fig:simplex} indicates, the non-linear one based on Ineq.~(\ref{eq:action_optimal_llr}) leads to a lower decision cost, particularly, a lower false acceptance rate.
Note that, Ineqs.~(\ref{eq:action_fused_lr}) and (\ref{eq:action_optimal_llr}) are agnostic to the  data distributions, and we use Gaussians in simulation only for convenience.

\begin{figure*}[t]
\resizebox{1.05\textwidth}{!}
{
        \centering
      \begin{subfigure}[t]{0.5\textwidth}
        \includegraphics[width=\textwidth]{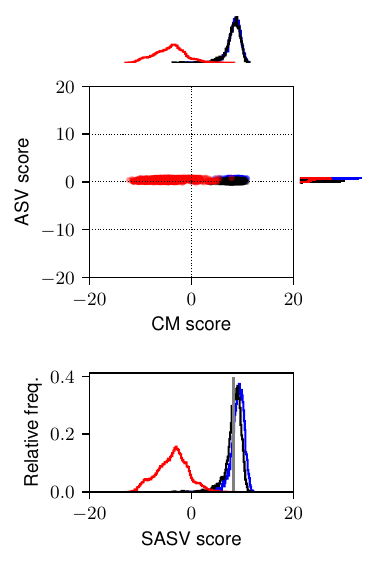}
        \vspace{-6mm}
        \caption{\sysbase{}}
        \label{fig:scatter:sub1}
     \end{subfigure}
     \hspace{-4mm}
      \begin{subfigure}[t]{0.5\textwidth}
        \includegraphics[width=\textwidth]{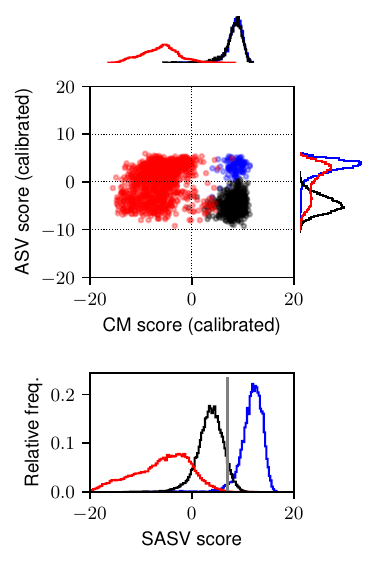}
        \vspace{-6mm}
        \caption{\sysbasec{}}
        \label{fig:scatter:sub2}
     \end{subfigure}
      \hspace{-5mm}
      \begin{subfigure}[t]{0.5\textwidth}
       \includegraphics[width=\textwidth]{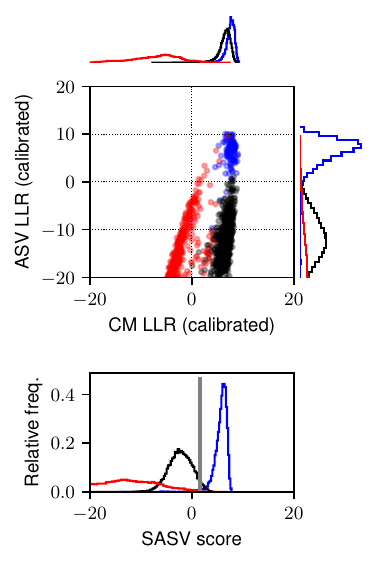}
       \vspace{-6mm}
       \caption{\syssic{}}
       \label{fig:scatter:sub3}
     \end{subfigure}
      \hspace{-4mm}
      \begin{subfigure}[t]{0.5\textwidth}
        \includegraphics[width=\textwidth]{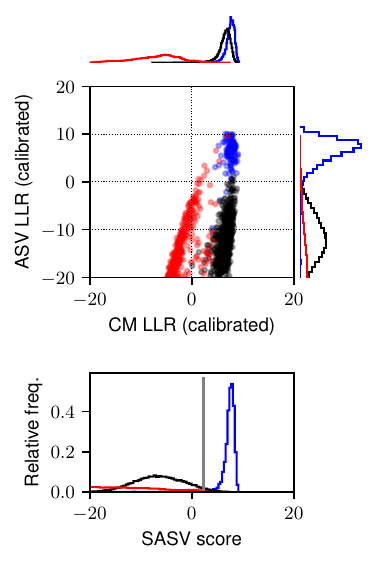}
        \vspace{-6mm}
        \caption{\syssiic{}}
        \label{fig:scatter:sub4}
     \end{subfigure}
     }
     \vspace{-3mm}
     \caption{Distributions of CM, ASV, and fused SASV scores. \textcolor{blue}{Bona fide data of target speakers ($\classtar{}$)},  \textcolor{black}{those of non-target speakers ($\classnon{}$)}, and \textcolor{red}{spoofed data ($\classspoof{}$)} are in different colors. Each vertical line in bottom plane marks SASV-EER threshold.
     }
     \vspace{-5mm}
     \label{fig:scatter}
\end{figure*}

On the basis of Ineq.~(\ref{eq:action_optimal_llr}), we define a new fusion method
\begin{equation}
\scoresasv = -\log\big[(1-\rho) e^{-\llr{\evidencenew{s}}{\classtar{}}{\classnon}} + \rho e^{- \llr{\evidencenew{s}}{\classtar{}}{\classspoof{}}}\big].
\label{eq:fusionthreegaussiansnl}
\end{equation}
The usage is similar to that in Sec.~\ref{sec:llrsum}, but we plug the LLRs into Eq.~(\ref{eq:fusionthreegaussiansnl}). 
Unfortunately, the prior-dependent $\rho$ cannot be decoupled from the LLRs, and it needs to be decided, for example, by grid search on a development set. The green solid line in Fig.~\ref{fig:simplex} shows the decision boundary when $\rho$ is computed from different priors. It performs reasonably well.

As a special case, if each $p(\evidencenew{s}; \boldsymbol{\theta} )$ is implemented as a Gaussian distribution, Eq.~(\ref{eq:fusionthreegaussiansnl}) turns out to be equivalent to the so-called Gaussian back-end fusion \cite[Eq.(1)]{todisco18_interspeech}. The link between Gaussian back-end fusion and the optimal decision has not been shown before as far as the authors are aware. Detailed proof is available on the code site (see link on page 1).

\section{Experiments}
\label{sec:exp}
We verified the effectiveness of the three fusion methods on the SASV challenge dataset \cite{jung22c_interspeech} and the official training, development, and evaluation splits.

\subsection{Model recipe and evaluation metrics}
All eight experimental systems were based on the open-sourced SASV challenge baseline B1 (Fig.~\ref{fig:sasv}) and used the pre-trained 
ECAPA-TDNN-based ASV \cite{desplanques20_interspeech} and AASIST-based CM \cite{jung2022aasist} embedding extractors, which were provided by the challenge organizers. The differences in the systems are
\begin{itemize}
\item \sysbase{} is a replicate of SASV B1. The ASV branch requires no training, and the neural network in the CM branch is trained using binary cross-entropy given $\scorecm$ and the labels for CM. The SASV score is computed by $\scoresasv = \scorecm + \scoreasv$. 
\item \syssi{} is similar to \sysbase{} but uses Eq.~(\ref{eq:fusionthreegaussians}) for score fusion. Each $p(\evidencenew{s};  \boldsymbol{\theta})$ is a Gaussian with a full covariance matrix. 
\item \syssii{} follows \syssi{} but uses Eq.~(\ref{eq:fusionthreegaussiansnl}) for score fusion.
\item \sysbasec{}, \syssic{}, and \syssiic{} are variants of \sysbase{}, \syssi{}, and \syssii{}, respectively, but calibrated scores or LLRs before fusion. The calibration function is $f(x)=ax+b$.
\item \syssrefi{} (B2-v2 in \cite{jung22c_interspeech}) and \syssrefii{} (PR-S-F in \cite{zhangProbabilistic2022}) were included for reference. They interpret scores as posterior probabilities and fuse by $\scoresasv = \sigma(\scorecm) +\scoreasv$ and $\scoresasv = \sigma(\scorecm)\times\sigma(\scoreasv)$, respectively \cite{kittlerCombining1998}. The $\sigma(\cdot)$ is a sigmoid function.
\end{itemize}
The parameters $\boldsymbol{\theta}$, $\rho$, and $\{a,b\}$ for calibration were tuned on the development set. 

All systems have about 82.5 k trainable parameters in the the neural network for CM, which has three hidden layers with 258, 128, and 64 neurons, respectively. All layers used a leaky ReLU activation with a negative slope value of 0.3.
All systems were trained on a Tesla V100 card using the Adam optimizer with a learning rate of $5\times 10^{-5}$, batch size of 24, and weight decay with a coefficient of $1\times 10^{-3}$. The maximum number of training epochs was 10, and the checkpoint with the best CM EER on the development set was used for evaluation. The implementation was done using SpeechBrain \cite{speechbrain}, and the training took less than one day per system.

The SASV-EER \cite{jung22c_interspeech} was used to gauge the discrimination power of $\scoresasv$. 
The log-likelihood-ratio cost $\cllr$ \cite{brummer2010measuring} was also computed and decomposed into $\cllr_{\text{min}}$ and $\cllr_{\text{calib}}$ to measure the discrimination and calibration performance of $\scoresasv$. 
The t-EER \cite{kinnunenTEER2023} was computed given $\{\scoreasv, \scorecm\}$ or the LLRs. 
Each system was trained and evaluated for six rounds, where each round used a different seed to randomly initialize the neural network in the CM sub-system. The  averaged results are listed in Table~\ref{tab}. The confidence interval of SASV-EER was using the Interspeech official toolkit with the default setting.

\input{tab1.tex}

\subsection{Results}
\noindent \textbf{Calibration before fusion was useful}: 
As reported in the original paper \cite{jung22c_interspeech}, \sysbase{} achieved an SASV-EER of around 20\%. As Fig.~\ref{fig:scatter}(a) shows, due to the large difference in the dynamic ranges of ASV and CM scores, the latter dominated the fused SASV score. 
The values of $\cllr$ and $\cllr_{\text{calib}}$ were higher than 1 bit, indicating that the SASV scores were not useful for making decisions.
By calibrating $\scoreasv$ and $\scorecm$ before fusion,  \sysbasec{} reduced the SASV EER to 2.73\%, and $\cllr$-based metrics also improved. From Fig.~\ref{fig:scatter}(b), we can observe that the distribution of $\scoreasv$ was `stretched out', and the fused $\scoresasv$ better separated the three classes of data. 

When fusing LLRs, the system pairs \{\syssi{}, \syssic{}\} and \{\syssii{}, \syssiic{}\} showed that the gain of the calibration was diminishing. This is understandable because the dynamic range of ASV and CM scores are handled by the covariance matrix of $p(\evidencenew{s};\boldsymbol{\theta})$. However, calibrating before fusion was still helpful, especially in the case of linear fusion in Eq.~(\ref{eq:fusionthreegaussians}). Compared with \syssi{},  \syssic{}'s EER was reduced to 1.56\%, and its $\cllr$ was only 0.14 bits. 

\noindent \textbf{Non-linear fusion of LLRs was better}: 
Rather than fusing the ASV and CM scores, \syssi{}, \syssii{}, and their variants fused the estimated LLRs. 
The nonlinear fusion of LLRs in \syssiic{} led to the best SASV-EER in our experiments. 
The difference between \syssiic{} and \syssic{} was small, possibly because the test data set of the SASV challenge can be well separated using a linear decision boundary in the space of LLRs, as Fig.~\ref{fig:scatter}(c) and (d) show. 

Compared with \syssrefi{} and \syssrefii{}, which treat the ASV and CM scores as posterior probabilities, \syssic{}, \syssii{}, and \syssiic{} performed similarly or better in terms of SASV-EER, and their $\cllr_{\text{calib}}$s were much lower. The results suggest that the fusion methods based on \ideaname{} are also practically useful.

Finally, the t-EER measures the discrimination power of $\{\scoreasv, \scorecm\}$ or LLRs and is agnostic to their calibration and fusion. Hence, the t-EERs of \sysbase{} and \sysbasec{} are the same. Similarly, the four systems fusing LLRs have the same t-EER. 
It was observed that using the LLRs led to a lower t-EER than raw or sigmoid-transformed $\scoreasv$ and $\scorecm$. This is evidence to encourage the sub-systems to produce LLRs.

\section{Conclusions}
This paper presented an alternative interpretation of SASV score fusion based on \ideaname{}. 
We suggested practices and improved fusion methods that may have been ignored or used ad hoc. First, the calibration of scores or LLRs before fusion was found to be helpful, especially when linearly fusing raw scores or LLRs. Second, rather than fusing raw scores, estimating and fusing LLRs turned out to be better in experiments. Third, the linear fusion of LLRs was found to be inferior to the non-linear one in both analysis and experiments. 
What is presented in this paper is by no means the only way to interpret SASV fusion, thus is expected to foster other studies.

\newpage
\section{Acknowledgements}
This work was supported by JST, PRESTO Grant Number JPMJPR23P9, Japan. It was supported in part by the Academy of Finland under Grant 349605, project "SPEECHFAKES".

\pretolerance=1000
\tolerance=200 
\emergencystretch=5pt

\bibliographystyle{IEEEtran}
\bibliography{mybib}

\newpage
\input{app}

\end{document}

%% file: note.tex
\onecolumn
{\noindent\Large \textbf{ISCA Copyright Notice}}

${}$

{\noindent\large \copyright 2024 ISCA. 
By each submission of a paper to INTERSPEECH-2024, the author(s) hereby warrant(s) that the manuscript has been submitted to ISCA (the International Speech Communication Association) for publication in the INTERSPEECH 2024 Proceedings, IS ORIGINAL AND HAS NOT BEEN SUBMITTED FOR PUBLICATION OR PUBLISHED ELSEWHERE and that all trademark use within the manuscript has been credited to its owner or written permission to use the name has been granted. In addition, the author(s) acknowledge(s) that all images such as tables, screenshots, graphics, etc., do not have a copyright that is held by a third party. The author(s) agree(s) to, and hereby assign(s) all rights, title and interest, including copyrights, in and to the manuscript to ISCA. The author(s) retain(s) the rights to any intellectual property developed by the author(s) and included in the manuscript. Note that the ISCA reviewing process is confidential. The copyright is provisionally transferred to ISCA at the time of submission of the article. This copyright is definitively assigned to ISCA if the manuscript is accepted after the reviewing process. If the manuscript is not accepted, the copyright transfer to ISCA is thus abrogated.

${}$

\noindent
This work is accepted by the ISCA INTERSPEECH 2024.

${}$

\noindent
\begin{verbatim}
@inproceedings{wangRevisiting2024,
  title = {Revisiting and Improving Scoring Fusion for 
       Spoofing-aware Speaker Verification Using Compositional 
       Data Analysis},
  booktitle = {Proc. {{Interspeech}}},
  author = {Wang, Xin and Kinnunen, Tomi and Kong Aik, Lee and 
      Noé, Paul-Gauthier and Yamagishi, Junichi},
  year = {2024},
  pages = {(accepted)}
}
\end{verbatim}
}
\twocolumn

%% file: tab1.tex

\setlength{\tabcolsep}{3pt}

\begin{table}[!t]
\caption{Results on SASV evaluation set. Each number is averaged over six training and evaluation rounds. Calibration (discriminative) is not on $\scoresasv$ but on scores or LLRs to be fused. Darker cell color indicates worse result in each row.}
\vspace{-6mm}
\begin{center}
\setlength{\tabcolsep}{2pt}
\resizebox{\columnwidth}{!}
{
\begin{tabular}{rrrrrrr||rr}
\toprule
  ID &   \sysbase{}   & \sysbasec{}  & \syssi{}  & \syssic{}  & \syssii{} & \syssiic{}  & \syssrefi{}  & \syssrefii{} \\ 
 \midrule
 Fusion & \multicolumn{2}{r}{Eq.~(\ref{eq:fusion_score})} & \multicolumn{2}{r}{Eq.~(\ref{eq:fusionthreegaussians})} & \multicolumn{2}{r||}{Eq.~(\ref{eq:fusionthreegaussiansnl})} & \cite{jung22c_interspeech} & \cite{zhangProbabilistic2022} \\
\cmidrule(lr){2-3}   \cmidrule(lr){4-5}  \cmidrule(lr){6-7} 
Calibration & $\times$ & $\checkmark$ & $\times$ & $\checkmark$  & $\times$ & $\checkmark$ & $\times$ & $\times$ \\
\midrule
      SASV-EER (\%)              & \cellcolor[rgb]{0.81, 0.81, 0.81} 20.46 & \cellcolor[rgb]{0.99, 0.99, 0.99} 2.73 & \cellcolor[rgb]{0.98, 0.98, 0.98} 3.31 & \cellcolor[rgb]{1.00, 1.00, 1.00} 1.56 & \cellcolor[rgb]{1.00, 1.00, 1.00} 1.44 & \cellcolor[rgb]{1.00, 1.00, 1.00} 1.43 & \cellcolor[rgb]{1.00, 1.00, 1.00} 1.60 & \cellcolor[rgb]{1.00, 1.00, 1.00} 1.55\\ 
      conf. ($\alpha=5\%$) & $\pm0.40$ & $\pm0.27$ & $\pm0.31$ & $\pm0.23$ & $\pm0.23$ & $\pm0.23$ & $\pm0.22$ & $\pm0.24$ \\
      \midrule
       $\mathrm{Cllr}$         & \cellcolor[rgb]{0.81, 0.81, 0.81} 2.17 & \cellcolor[rgb]{0.92, 0.92, 0.92} 1.09 & \cellcolor[rgb]{0.92, 0.92, 0.92} 1.04 & \cellcolor[rgb]{1.00, 1.00, 1.00} 0.14 & \cellcolor[rgb]{1.00, 1.00, 1.00} 0.18 & \cellcolor[rgb]{1.00, 1.00, 1.00} 0.16 & \cellcolor[rgb]{0.93, 0.93, 0.93} 0.96 & \cellcolor[rgb]{0.94, 0.94, 0.94} 0.84\\ 
 $\mathrm{Cllr}_{\text{min}}$  & \cellcolor[rgb]{0.81, 0.81, 0.81} 0.52 & \cellcolor[rgb]{0.98, 0.98, 0.98} 0.11 & \cellcolor[rgb]{0.97, 0.97, 0.97} 0.13 & \cellcolor[rgb]{1.00, 1.00, 1.00} 0.07 & \cellcolor[rgb]{1.00, 1.00, 1.00} 0.06 & \cellcolor[rgb]{1.00, 1.00, 1.00} 0.07 & \cellcolor[rgb]{0.99, 0.99, 0.99} 0.08 & \cellcolor[rgb]{1.00, 1.00, 1.00} 0.07\\ 
$\mathrm{Cllr}_{\text{calib}}$ & \cellcolor[rgb]{0.81, 0.81, 0.81} 1.64 & \cellcolor[rgb]{0.89, 0.89, 0.89} 0.98 & \cellcolor[rgb]{0.90, 0.90, 0.90} 0.91 & \cellcolor[rgb]{1.00, 1.00, 1.00} 0.07 & \cellcolor[rgb]{1.00, 1.00, 1.00} 0.11 & \cellcolor[rgb]{1.00, 1.00, 1.00} 0.10 & \cellcolor[rgb]{0.91, 0.91, 0.91} 0.88 & \cellcolor[rgb]{0.92, 0.92, 0.92} 0.78\\ 
 \midrule
  t-EER   (\%)           & \cellcolor[rgb]{0.85, 0.85, 0.85} 2.10 & \cellcolor[rgb]{0.85, 0.85, 0.85} 2.10 & \cellcolor[rgb]{1.00, 1.00, 1.00} 1.68 & \cellcolor[rgb]{1.00, 1.00, 1.00} 1.68 & \cellcolor[rgb]{1.00, 1.00, 1.00} 1.68 & \cellcolor[rgb]{1.00, 1.00, 1.00} 1.68 & \cellcolor[rgb]{0.81, 0.81, 0.81} 2.19 & \cellcolor[rgb]{0.81, 0.81, 0.81} 2.21\\ 
\bottomrule
\end{tabular}
}
\vspace{-9mm}
\end{center}
\label{tab}
\end{table}%

%% file: app.tex
\newpage
\clearpage
\appendix

\newcommand{\costmiss}{C_{\text{miss}}^{\classtar{}}}
\newcommand{\costfanon}{C_{\text{fa}}^{\classnon{}}}
\newcommand{\costfaspf}{C_{\text{fa}}^{\classspoof{}}}

\copyright 2024 Xin Wang 

\section{Appendix}

\subsection{Optimal decision policy}
\label{seq:app_opt}
We elaborate on the decision policy and linear and non-linear decision boundaries presented in Sec.~\ref{sec:nonlinear}. 

\subsubsection{SASV hypotheses and actions}
In SASV, or the joint operation of ASV and CM, the input pair of the probe and reference $\evidence =(\evidence^{(p)}, \evidence^{(r)}) \in\mathcal{X}$, either as waveforms or feature vectors, can be from one of the three possible classes $\mathcal{Y}=\{\symbolhp{}_\classtar{}, \symbolhp{}_\classnon{}, \symbolhp{}_\classspoof{} \}$\cite{hadid2015biometrics}:
\begin{align*} 
\classtar{}:  & \, \evidence_{p} \text{ is bona fide \emph{and} matches the speaker identity in }  \evidence_{r}, \\
\classnon{}: & \,\evidence_{p} \text{ is bona fide \emph{but} the speaker identity does not match},\\
\classspoof{}: &\, \evidence_{p} \text{ is spoofed}.
\end{align*} 
In the last class, the SASV system does not need to differentiate whether a spoofed probe $\evidence_{p}$ sounds like the target speaker. A more detailed discussion is given in Sec.~\ref{sec:app_spoofed}.

SASV is used to accept only bona fide speech uttered by the target speaker. Therefore, an SASV system can take one of two possible actions $\mathcal{A} = \{\mathtt{accept}, \mathtt{reject}\}$:
\begin{align*}
\mathtt{accept}: &\, \evidence_{p} \text{ is claimed to be from } \classtar{} \text{ and accepted},\\
\mathtt{reject}: &\, \evidence_{p} \text{ is claimed to be NOT from } \classtar{} \text{ and rejected}.
\end{align*}
In the case of $\mathtt{reject}$, SASV does not need to differentiate whether $\evidence_{p}$ is $\classnon{}$ or $\classspoof{}$.

\subsubsection{Decision cost and optimal decision policy}
An SASV system may take wrong actions. A sample from $\classtar{}$ may be falsely rejected (missed), while samples from $\classnon{}$ or $\classspoof{}$ may be falsely accepted. 
We can assign a decision cost to each incorrect decision.
Given an action $a\in\mathcal{A}$ and the ground-truth label $y\in\mathcal{Y}$, we use $C(a, y)\geq0$ to denote the decision-cost function $c: \mathcal{A}\times{\mathcal{Y}}\rightarrow\mathbb{R}^{+}$.
A general definition of $C(a, y)$ is given in Table~\ref{tab:app_cost}.  

Following the Bayesian decision theory \cite{dudaPattern2001}, we can compute the conditional risk $R(a | \evidence)$ when taking an action $a$ given the input sample $\evidence$ as
\begin{equation}
R(a | \evidence) = \sum_{y\in\mathcal{Y}}C(a, y) P(y | \evidence), 
\end{equation}
where $P(y | \evidence)$ is the posterior probability. 
By plugging in the possible values of $C(a, y)$ listed in Table~\ref{tab:app_cost}, we can compute the conditional risks when taking the two possible actions
\begin{align}
R(\mathtt{accept} | \evidence) &= \costfanon{} P(\symbolhp{}_\classnon{} | \evidence) + \costfaspf{} P(\symbolhp{}_\classspoof{} | \evidence), \label{eq:app_cost1} \\ 
R(\mathtt{reject} | \evidence) &= \costmiss{} P(\symbolhp{}_\classtar{} | \evidence). \label{eq:app_cost2}
\end{align}

To minimize the overall risk, the Bayes decision rule tells us to select the action with a smaller conditional risk \cite{dudaPattern2001}. Accordingly, the decision policy is
\begin{equation}
{a} = 
\begin{cases}
\mathtt{accept}, \quad \text{if } R(\mathtt{reject} | \evidence) > R(\mathtt{accept} | \evidence) \\
\mathtt{reject}, \quad \text{if } R(\mathtt{reject} | \evidence)  \leq R(\mathtt{accept} | \evidence) \\
\end{cases}.
\end{equation}
In other words, the optimal decision policy for an SASV system is to $\mathtt{accept}$ the input probe if the conditional risk $R(\mathtt{accept} | \evidence)$ is smaller than $R(\mathtt{reject} | \evidence)$. Otherwise, the system should $\mathtt{reject}$ the input.

\begin{table}[!t]
\caption{Values of decision cost function $C(a, y)$ for SASV. Subscript fa means false acceptance. }
\vspace{-5mm}
\begin{center}
\begin{tabular}{rcll}
\toprule
& & \multicolumn{2}{c}{Action $a$} \\
\cmidrule{3-4}
True class label $y$ && $\mathtt{accept}$ & $\mathtt{reject}$\\
\midrule
$\classtar{}$  && 0 & $\costmiss$ \\
$\classnon{}$ && $\costfanon$ & 0\\
$\classspoof{}$ && $\costfaspf$ & 0\\
\bottomrule
\end{tabular}
\end{center}
\label{tab:app_cost}
\end{table}%

With Eqs.~(\ref{eq:app_cost1}) and (\ref{eq:app_cost2}), we re-write the condition to accept the input, i.e., $R(\mathtt{reject} | \evidence) > R(\mathtt{accept} | \evidence)$, as 
\begin{equation}
\begin{split}
\costmiss{} P(\symbolhp{}_\classtar{} | \evidence) &>  \\
\costfanon{} & P(\symbolhp{}_\classnon{} | \evidence)  + \costfaspf{} P(\symbolhp{}_\classspoof{} | \evidence).
\label{eq:app_policy_gen}
\end{split}
\end{equation}
Using the Bayes' formula, we write the left and right hand sides as 
\begin{equation}
\costmiss{} P(\symbolhp{}_\classtar{} | \evidence) = \costmiss{}\frac{ p(\evidence | \symbolhp{}_\classtar{}) \pi_{\classtar{}}}{p(\evidence)},
\end{equation}
and
\begin{equation}
\begin{split}
 & \costfanon{} P(\symbolhp{}_\classnon{} | \evidence)  + \costfaspf{} P(\symbolhp{}_\classspoof{} | \evidence) \\
=  & \costfanon{} \frac{p(\evidence | \symbolhp{}_\classnon{} ) \pi_{\classnon{}}}{p(\evidence)} + \costfaspf{} \frac{p( \evidence | \symbolhp{}_\classspoof{} ) \pi_{\classspoof{}}}{p(\evidence)}.
\end{split}
\end{equation}
Plugging the above two equations into Ineq.~(\ref{eq:app_policy_gen}) and cancel $p(\evidence)$ on both sides of the inequality, Ineq.~(\ref{eq:app_policy_gen}) can be written in terms of likelihoods and priors as
\begin{equation}
\begin{split}
 \costmiss{} p(\evidence | \symbolhp{}_\classtar{})  \pi_{\classtar{}} &> \\
  \costfanon{} p(\evidence | \symbolhp{}_\classnon{} ) &  \pi_{\classnon{}} + \costfaspf{} p( \evidence | \symbolhp{}_\classspoof{} ) \pi_{\classspoof{}}.
\label{eq:app_policy_gen_llr}
\end{split}
\end{equation}
By dividing both sides by $\costmiss{} p(\evidence | \symbolhp{}_\classtar{})$, we obtain
\begin{equation}
\begin{split}
 \pi_{\classtar{}} > 
  \frac{\costfanon{} p(\evidence | \symbolhp{}_\classnon{} )}{\costmiss{} p(\evidence | \symbolhp{}_\classtar{})  } \pi_{\classnon{}} + \frac{\costfaspf{} p( \evidence | \symbolhp{}_\classspoof{} ) }{\costmiss{} p(\evidence | \symbolhp{}_\classtar{}) } \pi_{\classspoof{}},
\label{eq:app_policy_gen_llr2}
\end{split}
\end{equation}
which can be further written as 
\begin{equation}
\begin{split}
 \pi_{\classtar{}} > 
  \frac{\costfanon{}}{\costmiss{}} e^{-\llr{\evidence}{\classtar{}}{\classnon{}}} \pi_{\classnon{}} + \frac{\costfaspf{} }{\costmiss{} } e^{-\llr{\evidence}{\classtar{}}{\classspoof{}}} \pi_{\classspoof{}},
\label{eq:app_policy_gen_llr3}
\end{split}
\end{equation}
where the LLRs are defined as $ \llr{\evidence}{\classtar{}}{\classnon{}} \triangleq \log\frac{p( \evidence | \symbolhp{}_{\classtar{}})}{p( \evidence | \symbolhp{}_{\classnon{}})}$ and $ \llr{\evidence}{\classtar{}}{\classspoof{}} \triangleq \log\frac{p( \evidence | \symbolhp{}_{\classtar{}})}{p( \evidence | \symbolhp{}_{\classspoof{}})}$.

Because the samples from $\classnon{}$ and $\classspoof{}$ should be rejected, they can be conceptually grouped together as a ``negative'' class.  The prior probability of having a sample from this ``negative'' class is therefore $\pi_{\classnon{}} + \pi_{\classspoof{}}$, and, among the ``negative'' samples,  the proportion of $\pi_{\classspoof}$ is defined as
\begin{equation}
\rho \triangleq \frac{ \pi_{\classspoof{}} }{\pi_{\classnon{}} + \pi_{\classspoof{}}}.
\end{equation}
The ratio $\rho$ is referred to as \emph{spoof prevalence prior} in the t-EER \cite{kinnunenTEER2023}. Furthermore, The ratio of the priors of the $\classtar{}$ against the ``negative class'' can be defined as
\begin{equation}
\beta \triangleq \frac{ \pi_{\classtar{}} }{\pi_{\classnon{}} + \pi_{\classspoof{}}} = \frac{ \pi_{\classtar{}} }{1-\pi_{\classtar{}}}.
\end{equation}

By dividing both sides of Ineq.~(\ref{eq:app_policy_gen_llr3}) with $\pi_{\classnon{}} + \pi_{\classspoof{}}$, we finally obtain 
\begin{equation}
\begin{split}
\beta > 
  \frac{\costfanon{}}{\costmiss{}} e^{-\llr{\evidence}{\classtar{}}{\classnon{}}} (1-\rho) + \frac{\costfaspf{} }{\costmiss{} } e^{-\llr{\evidence}{\classtar{}}{\classspoof{}}} \rho,
\label{eq:app_policy_gen_llr4}
\end{split}
\end{equation}

In short, to make the optimal decision, an SASV system computes the right hand side of Ineq.~(\ref{eq:app_policy_gen_llr4}) given the LLRs and the priors. The value can be interpreted as a `score'. The system then compares the score with the left hand side to make a decision.  If the score is smaller than $\beta$, the system $\mathtt{accept}$ the input. 

Note that the procedure assumes that a lower score indicates favors $\mathtt{accept}$ more. In most implementations, it is more intuitive if a \emph{higher} score indicates favors $\mathtt{accept}$ more. This can be easily done re-written Ineq.(\ref{eq:app_policy_gen_llr4}), for example, as
\begin{equation}
\begin{split}
-\Big[\frac{\costfanon{}}{\costmiss{}} e^{-\llr{\evidence}{\classtar{}}{\classnon{}}} (1-\rho) + \frac{\costfaspf{} }{\costmiss{} } e^{-\llr{\evidence}{\classtar{}}{\classspoof{}}} \rho\Big]
> -\beta.
\end{split}
\end{equation}
We can further take $\log(\cdot)$ on both sides. Notice that the left hand side increases if the two LLRs increase. In other words, if the LLRs give more weights to $\classtar{}$, the decision favors $\mathtt{accept}$ more.

\subsubsection{On decoupling LLR and priors}
Let us consider an extreme case without spoofed data, i.e., $\rho=0$ and $\pi_\classspoof{} = 0$. 
Ineq.~(\ref{eq:app_policy_gen_llr4}) can be written as  
\begin{equation}
\begin{split}
\beta >  \frac{\costfanon{}}{\costmiss{}} e^{-\llr{\evidence}{\classtar{}}{\classnon{}}},
\label{eq:app_policy_asv_llr}
\end{split}
\end{equation}
or, with $-\log(\cdot)$ applied to both sides of the inequality, 
\begin{equation}
\begin{split}
\llr{\evidence}{\classtar{}}{\classnon{}} > \log \frac{1-\pi_\classtar{}}{\pi_\classtar{}} \frac{\costfanon{}}{\costmiss{}}.
\label{eq:app_policy_asv_llr2}
\end{split}
\end{equation}
This is the optimal decision policy for ASV as a typical binary classification task \cite[(2.33)]{nautsch2019speaker}. 

Note that the LLRs are on the left hand side, while the priors (and decision costs) are on the right hand side.  
Because the LLRs are decoupled from the priors, the system can be designed to produce the LLRs without considering the priors. It is the user, who deploys the system into an application, assign the decision costs, asserted priors, and accordingly the decision threshold. 
The decoupled LLRs and priors are preferred in a forensic context \cite{castro2007forensic, nautsch2019speaker}.  A similar decision policy can be derived when $\rho=1$, i.e., CM without ASV.

Unfortunately, when $\rho\in(0, 1)$, the LLRs and priors cannot be decoupled. The $\rho$ is needed to compute the score, and the SASV system designer has to estimate the value of the $\rho$, for example, on the basis of the statistics from the development data. In our experiment, $\rho$ was found through grid search by minimizing the decision risks on the development data. 

\subsection{Optimal decision policy when decision costs are equal}

For an actual application, we need to assign values to the decision costs on the basis of the application's requirements.  For example, the cost for missing a sample of $\classtar{}$ may be high for a smart-speech device since the user may be annoyed if their voice is falsely rejected. However, the costs for falsely accepting $\classspoof{}$ and $\classnon{}$ may be high for online banking. False acceptance of imposters may lead to disastrous financial loss. 

Following the common assumptions in other studies (e.g., \cite{todisco18_interspeech}), we now consider a particular case in which $\costfanon{} = \costfaspf{} = \costmiss{}$. 
Accordingly, the optimal policy of action for the SASV system is to $\mathtt{accept}$ when
\begin{equation}
\begin{split}
P(\symbolhp{}_\classtar{} | \evidence) >  P(\symbolhp{}_\classnon{} | \evidence)  + P(\symbolhp{}_\classspoof{} | \evidence),
\label{eq:app_policy}
\end{split}
\end{equation}
or, equivalently, 
\begin{equation}
\begin{split}
\beta > 
  e^{-\llr{\evidence}{\classtar{}}{\classnon{}}} (1-\rho) + e^{-\llr{\evidence}{\classtar{}}{\classspoof{}}} \rho.
\label{eq:app_policy_llr4}
\end{split}
\end{equation}
As mentioned above, the SASV system needs to compute the right hand side of Ineq.~(\ref{eq:app_policy_llr4}) as a score. The score is then compared with $\beta$ to make a decision. 

In practice, we assume that a higher SASV score indicates that the input is more likely from $\classtar{}$. Hence, we take the negative logarithm over the two sides of Ineq.~(\ref{eq:app_policy_llr4}) and obtain 
\begin{equation}
\begin{split}
-\log\big[e^{-\llr{\evidence}{\classtar{}}{\classnon{}}} (1-\rho) + e^{-\llr{\evidence}{\classtar{}}{\classspoof{}}} \rho\big] > -\log \beta.
\label{eq:app_policy_llr5}
\end{split}
\end{equation}
Taking the logarithm makes the left hand side similar to conducting a log-sum-exp operation on the LLRs, \footnote{https://en.wikipedia.org/wiki/LogSumExp} albeit a weighted summation with weights $\rho$ and $1-\rho$. 

In implementation, the SASV system computes the score $\scoresasv = -\log\big[e^{-\llr{\evidence}{\classtar{}}{\classnon{}}} (1-\rho) + e^{-\llr{\evidence}{\classtar{}}{\classspoof{}}} \rho\big]$ for an input $\evidence$. It compares $\scoresasv$ with the threshold $-\log\beta$ for making the decision. 

Recall that if we assume the ASV and CM score vector $\evidencenew{s}=[\scoreasv, \scorecm]^{\top}$ encodes all the relevant information about the input sample $\evidence$, we can compute the LLRs given $\evidencenew{s}$ as the observed data. We then obtain the scoring function in Eq.~(\ref{eq:fusionthreegaussiansnl}) of Sec.~\ref{sec:nonlinear}.

\subsubsection{Link to Gaussian fusion back-end}
Note that 
\begin{equation}
\begin{split}
\scoresasv &= -\log\big[e^{-\llr{\evidence}{\classtar{}}{\classnon{}}} (1-\rho) + e^{-\llr{\evidence}{\classtar{}}{\classspoof{}}} \rho\big] \\
&= -\log\Big[\frac{p( \evidence | \symbolhp{}_{\classnon{}})}{p( \evidence | \symbolhp{}_{\classtar{}})} (1-\rho) + 
\frac{p( \evidence | \symbolhp{}_{\classspoof{}})}{p( \evidence | \symbolhp{}_{\classtar{}})} \rho\Big] \\
&= -\log\Big[\frac{(1-\rho) p( \evidence | \symbolhp{}_{\classnon{}}) + \rho p( \evidence | \symbolhp{}_{\classspoof{}})}{p( \evidence | \symbolhp{}_{\classtar{}})} \Big] \\
&= \log\frac{p( \evidence | \symbolhp{}_{\classtar{}})}{(1-\rho) p( \evidence | \symbolhp{}_{\classnon{}}) + \rho p( \evidence | \symbolhp{}_{\classspoof{}})}.
\label{eq:app_gaussian}
\end{split}
\end{equation}
This equation has the same form as the so-called Gaussian fusion back-end \cite[Eq.(1)]{todisco18_interspeech}, but we have shown how it is derived from the optimal decision policy, given the decision-cost matrix defined in Table~\ref{tab:app_cost} and configuration $\costfanon{} = \costfaspf{} = \costmiss{}$.

Note that the Gaussian fusion back-end parameterizes each likelihood function using a Gaussian distribution, but any other parametric distribution can be used.

\subsubsection{Link to the sub-optimal decision policy}
In Sec.~\ref{sec:nonlinear}, we showed that the linear fusion of LLRs is linked to the decision policy in Ineqs.~(\ref{eq:action_fused}) and (\ref{eq:action_fused_lr}). For convenience, we write Ineqs.~(\ref{eq:action_fused}) again below
\begin{equation}
P(\symbolhp{}_{\classtar{}} | \evidence  ) ^ 2 > P(\symbolhp{}_{\classnon{}} | \evidence  )P(\symbolhp{}_{\classspoof{}} | \evidence),
\label{eq:app_linear_policy}
\end{equation} 

It has been described that any vector of posterior probabilities $\symbolvec{P}= [P(\symbolhp{}_{\classspoof{}} | \evidence),  P(\symbolhp{}_{\classnon{}} | \evidence  ), P(\symbolhp{}_{\classtar{}} | \evidence  )]^\top$ that satisfies Ineq.~(\ref{eq:app_policy}) also satisfies Ineq.~(\ref{eq:app_linear_policy}), but the reverse is not always true.  In other words, Ineq.~(\ref{eq:app_linear_policy}) is a necessary but not sufficient condition to achieve Ineq.~(\ref{eq:app_policy}). 

\noindent \textbf{Ineq.~(\ref{eq:app_linear_policy}) is necessary to achieve Ineq.~(\ref{eq:app_policy})}:
Note that $P(\symbolhp{}_{\classtar{}} | \evidence  )+ P(\symbolhp{}_{\classnon{}} | \evidence  ) + P(\symbolhp{}_{\classspoof{}} | \evidence) = 1$. Therefore, Ineq.~(\ref{eq:app_policy}) can be written as
\begin{equation}
\begin{split}
P(\symbolhp{}_\classtar{} | \evidence) &>  P(\symbolhp{}_\classnon{} | \evidence)  + P(\symbolhp{}_\classspoof{} | \evidence) \\
&= 1-  P(\symbolhp{}_\classtar{} | \evidence), 
\end{split}
\end{equation}
or, equivalently, $P(\symbolhp{}_\classtar{} | \evidence) > 0.5$. Therefore, for a $\symbolvec{P}$ that satisfies Ineq.~(\ref{eq:app_policy}), we know that $P(\symbolhp{}_\classtar{} | \evidence) > 0.5$, $P(\symbolhp{}_\classnon{} | \evidence) < 0.5$, $P(\symbolhp{}_\classspoof{} | \evidence) < 0.5$, and 
\begin{equation}
P(\symbolhp{}_\classtar{} | \evidence)^2 > 0.25 >P(\symbolhp{}_\classnon{} | \evidence)P(\symbolhp{}_\classspoof{} | \evidence).
\end{equation}
Hence, Ineq.~(\ref{eq:app_linear_policy}) is a necessary condition of Ineq.~(\ref{eq:app_policy}).

\noindent \textbf{Ineq.~(\ref{eq:app_linear_policy}) is not a sufficient condition of Ineq.~(\ref{eq:app_policy})}:
In Seq.~\ref{sec:nonlinear}, we provided an example  $\boldsymbol{P} = [0.05, 0.65, 0.3]$ that satisfies Ineq.~(\ref{eq:app_linear_policy}) but not Ineq.~(\ref{eq:app_policy}): $0.3^2 > 0.05 \times 0.65$ but $0.3 < 0.05 + 0.65$. Hence, Ineq.~(\ref{eq:app_linear_policy}) is not a sufficient condition of Ineq.~(\ref{eq:app_policy}).

\subsection{Comparing with ternary classification}
In Sec.\ref{seq:app_opt}, we defined SASV as a classification task with two actions but three data classes. We may compare it with a ternary classification task. For convenience, let the three classes be $\mathcal{Y}=\{\symbolhp{}_1, \symbolhp{}_2, \symbolhp{}_3\}$. The three possible actions are to assign the input to one of the three classes. 

\subsubsection{Minimizing the decision cost}
We can define $C(a, y)$ in Table~\ref{tab:app_cost_ter}.
According to the Bayes decision rule, the system should classify the input sample $\evidence$ to $\symbolhp{}_1$ if 
\begin{equation}
\sum_{i}C_{i1}P(\symbolhp{}_i | \evidence) < \sum_{i}C_{i2}P(\symbolhp{}_i | \evidence)
\end{equation}
\emph{and}
\begin{equation}
\sum_{i}C_{i1}P(\symbolhp{}_i | \evidence) < \sum_{i}C_{i3}P(\symbolhp{}_i | \evidence).
\end{equation}
Similar conditions can be written for accepting $\symbolhp{}_2$ and $\symbolhp_{3}$. 

If $C_{ii} = 0$, i.e., no cost for correct action, the condition is 
\begin{equation}
\begin{split}
C_{21}P(\symbolhp{}_2 | \evidence) + &C_{31}P(\symbolhp{}_3 | \evidence) \\
& < C_{12}P(\symbolhp{}_1 | \evidence) + C_{32}P(\symbolhp{}_3 | \evidence)
\end{split}
\end{equation}
\emph{and}
\begin{equation}
\begin{split}
C_{21}P(\symbolhp{}_2 | \evidence) + &C_{31}P(\symbolhp{}_3 | \evidence) \\
&< C_{23}p(\symbolhp{}_2 | \evidence) + C_{13}p(\symbolhp{}_1 | \evidence).
\end{split}
\end{equation}

If we further set $C_{32}=C_{23}=0$ as Table~\ref{tab:app_cost_ter2} shows, the condition becomes 
\begin{equation}
\begin{split}
C_{21}P(\symbolhp{}_2 | \evidence) + C_{31}P(\symbolhp{}_3 | \evidence)  < C_{12}P(\symbolhp{}_1 | \evidence)
\end{split}
\end{equation}
\emph{and}
\begin{equation}
\begin{split}
C_{21}P(\symbolhp{}_2 | \evidence) + C_{31}P(\symbolhp{}_3 | \evidence) < C_{13}p(\symbolhp{}_1 | \evidence).
\end{split}
\end{equation}

Merging the two sub-conditions, we further get 
\begin{equation}
\begin{split}
C_{21}P(\symbolhp{}_2 | \evidence) + C_{31}P(\symbolhp{}_3 | \evidence) < \min(C_{13}, C_{12})p(\symbolhp{}_1 | \evidence).
\end{split}
\end{equation}

\begin{table}[!t]
\caption{Values of decision cost function $C(a, y)$ for ternary classification. }
\vspace{-5mm}
\begin{center}
\begin{tabular}{rclll}
\toprule
& & \multicolumn{3}{c}{Action $a$} \\
\cmidrule{3-5}
True class label $y$ && 1 & 2 & 3\\
\midrule
1 & & $C_{11}$ & $C_{12}$ & $C_{13}$ \\
2 & & $C_{21}$ & $C_{22}$ & $C_{23}$ \\
3 & & $C_{31}$ & $C_{32}$ & $C_{33}$\\
\bottomrule
\end{tabular}
\end{center}
\label{tab:app_cost_ter}
\end{table}%

\begin{table}[!t]
\caption{Example values of decision cost function $C(a, y)$.}
\vspace{-5mm}
\begin{center}
\begin{tabular}{rclll}
\toprule
& & \multicolumn{3}{c}{Action $a$} \\
\cmidrule{3-5}
True class label $y$ && 1 & 2 & 3\\
\midrule
1 & & 0 & $C_{12}$ & $C_{13}$ \\
2 & & $C_{21}$ & 0 & $0$ \\
3 & & $C_{31}$ & 0 & 0\\
\bottomrule
\end{tabular}
\vspace{-5mm}
\end{center}
\label{tab:app_cost_ter2}
\end{table}%

\begin{table}[!t]
\caption{Values of decision utility function $U(a, y)$ for ternary classification task. }
\vspace{-5mm}
\begin{center}
\begin{tabular}{rclll}
\toprule
& & \multicolumn{3}{c}{Action $a$} \\
\cmidrule{3-5}
True class label $y$ && 1 & 2 & 3\\
\midrule
1 & & $U_{11}$ & $U_{12}$ & $U_{13}$ \\
2 & & $U_{21}$ & $U_{22}$ & $U_{23}$ \\
3 & & $U_{31}$ & $U_{32}$ & $U_{33}$\\
\bottomrule
\end{tabular}
\end{center}
\label{tab:app_utility_ter}
\end{table}%

\begin{table}[!t]
\caption{Example values of decision utility function $U(a, y)$.}
\vspace{-5mm}
\begin{center}
\begin{tabular}{rclll}
\toprule
& & \multicolumn{3}{c}{Action $a$} \\
\cmidrule{3-5}
True class label $y$ && 1 & 2 & 3\\
\midrule
1 & & $U$ & 0 & 0 \\
2 & & 0 & $U$ &  0 \\
3 & & 0 & 0 & $U$ \\
\bottomrule
\end{tabular}
\vspace{-5mm}
\end{center}
\label{tab:app_utility_ter_2}
\end{table}%

If we further assume $C_{12}=C_{13}$, we get
\begin{equation}
\begin{split}
C_{21}P(\symbolhp{}_2 | \evidence) + C_{31}P(\symbolhp{}_3 | \evidence) <  C_{12}p(\symbolhp{}_1 | \evidence),
\end{split}
\end{equation}
which has a similar form to the condition for SASV in Ineq.~(\ref{eq:app_policy_gen}). 

\subsubsection{Maximizing the decision utility}
In contrast to minimizing the decision cost, we may conduct a Bayesian decision by maximizing the utility \cite{he2006three}. Let us define a utility function $U(a, y)$ that returns the utility (or reward) when the system the action $a$ while the ground-truth label is $y$. 
A generic value table of $U(a, y)$ for a ternary classification task is shown in Table~\ref{tab:app_utility_ter}.  
Note that, different from the decision cost, we should set utilities so that $U_{ii} > U_{ij}\geq 0, i\neq{j}$.

The system accepts $\symbolhp{}_1$ if
\begin{equation}
\sum_{i}U_{i1}P(\symbolhp{}_i | \evidence) > \sum_{i}U_{i2}P(\symbolhp{}_i | \evidence)
\end{equation}
\emph{and}
\begin{equation}
\sum_{i}U_{i1}P(\symbolhp{}_i | \evidence) > \sum_{i}U_{i3}P(\symbolhp{}_i | \evidence).
\end{equation}
Similar conditions can be written for accepting $\symbolhp{}_2$ and $\symbolhp_{3}$. 

Although the condition looks similar to that for minimizing the decision cost, a nuance is that we can define a diagonal utility matrix, i.e., $U_{ii} > 0$ and $U_{ij}=0, i\neq{j}$. This means that only the correct decision receives a reward.
In this case, the condition to accept $\symbolhp{}_1$ can be written as
\begin{equation}
U_{11}P(\symbolhp{}_1 | \evidence) > U_{22}P(\symbolhp{}_2 | \evidence)
\end{equation}
\emph{and}
\begin{equation}
U_{11}P(\symbolhp{}_1 | \evidence) > U_{33}P(\symbolhp{}_3 | \evidence).
\end{equation}
If we further set $U_{ii} = U_{jj}=U, \forall{i, j}$ as Table~\ref{tab:app_utility_ter_2} shows, the condition becomes
\begin{equation}
P(\symbolhp{}_1 | \evidence) > P(\symbolhp{}_2 | \evidence) \land P(\symbolhp{}_1 | \evidence) > P(\symbolhp{}_3 | \evidence),
\label{eq:app_policy_utility}
\end{equation}
where $\land$ is the logical operator AND.

In \cite[sec.10.3]{noe:tel-04264175}, Ineq.~(\ref{eq:app_policy_utility}) and its relationship with \ideaname{} is discussed. 
Inequality.~(\ref{eq:app_policy_utility}) is different from Ineq.~(\ref{eq:app_policy}): the former is derived when maximizing the decision utility given a scalar utility matrix.

\subsection{On spoofed sample}
\label{sec:app_spoofed}
A spoofed sample should be rejected by SASV regardless of whether it sounds like the target speaker. If the spoofed probe is forged by an attacker (i.e., biometric imposters in ISO/IEC 30107), it is supposed to sound like the target speaker's voice; however, the algorithm to create the spoofed probe may not be good enough to mimic the voice. Hence, the spoofed sample should be rejected regardless of its similarity to the target speaker's voice. 

Suppose the `spoofed' sample is created by an authentic user who wants to make it less similar to their own voice (i.e., biometric concealers). In that case, this application is considered more relevant to voice anonymization \cite{tomashenkoIntroducingVoicePrivacyInitiative2020} rather than SASV. Of course, an SASV system should also reject the `spoofed' sample since it intended use is to conceal the speaker's identity.